# Investigating the Effect of Magnetic Dipole-Dipole Interaction on Magnetic Particle Spectroscopy (MPS): Implications for Magnetic Nanoparticle-based Bioassays and Magnetic Particle Imaging (MPI)

Kai Wu[†], Diqing Su[‡], Renata Saha[†], Jinming Liu[†], and Jian-Ping Wang[†,*]

[†]Department of Electrical and Computer Engineering, University of Minnesota, Minneapolis, Minnesota 55455, USA

[‡]Department of Chemical Engineering and Material Science, University of Minnesota, Minneapolis, Minnesota 55455, USA

*Corresponding author E-mail: jpwang@umn.edu

(Dated: January 4, 2019)

**Abstract**

Superparamagnetic iron oxide nanoparticles (SPIONs), with comparable size to biomolecules (such as proteins, nucleic acids, etc.) and unique magnetic properties, good biocompatibility, low toxicity, potent catalytic behavior, are promising candidates for many biomedical applications. There is one property present in most SPION systems, yet it has not been fully exploited, which is the dipole-dipole interaction (also called dipolar interaction) between the SPIONs. It is known that the magnetic dynamics of an ensemble of SPIONs are substantially influenced by the dipolar interactions. However, the exact way it affects the performance of magnetic particle-based bioassays and magnetic particle imaging (MPI) is still an open question. The purpose of this paper is to give a partial answer to this question. This is accomplished by numerical simulations on the dipolar interactions between two nearby SPIONs and experimental measurements on an ensemble of SPIONs using our lab-based magnetic particle spectroscopy (MPS) system. Our results show that even moderate changes in the SPION concentration may have substantial effects on the magnetic dynamics of the SPION system and the harmonic signal magnitudes can be increased or decreased by 60%, depending on the values of MPS system parameters.

1. Introduction

In the past decade, superparamagnetic iron oxide nanoparticles (SPIONs) and their liquid dispersions are the subject of intense research due to their unique magnetic and physical properties and their potential applications in several fields such as hyperthermia therapy[1-4], catalysis[1, 5-8], drug/gene delivery[9-14], magnetic bioassays[15-18], magnetic particle imaging (MPI) and magnetic resonance imaging (MRI)[1, 19-24]. Besides their unique magnetic properties, comparable size to biomolecules, low toxicity, and good biocompatibility, the recent advances in the area of chemical engineering and material science have enabled the facile synthesis and surface functionalization of SPIONs, placing them as one of the most popular nanocomposites in many biological and biomedical applications.

SPIONs show zero magnetization in the absence of magnetic fields, while an external magnetic field can magnetize the SPIONs, like a paramagnet but with larger magnetic susceptibility. Under an external magnetic field, these magnetized SPIONs exert dipolar fields (also called stray fields), which allows the detection and measurement of the SPIONs in magnetometers and susceptometers. On the other hand, the magnetic dynamics of an ensemble of SPIONs are known to be substantially influenced by the dipolar interactions[25-28]. This is due to the fact that, for a highly concentrated SPION suspension under an external magnetic field, the effective magnetic field on each SPION is reduced because of the dipolar fields.

In this work, we report a magnetic particle spectroscopy (MPS)-based method to investigate how magnetic dipolar interactions modify the responses of an ensemble of SPIONs in the presence of the external magnetic fields. At first, this investigation is carried out via numerical simulations based on the Object Oriented MicroMagnetic Framework (OOMMF) micromagnetic software[29, 30], followed by experimental measurements on different SPIONs suspensions with nanoparticle concentrations varying from 0.1133 to 3.4 nmole/mL (volume fractions varying from 0.001 to 0.029). It is seen that even moderate changes in the SPION concentration may have substantial effects on the magnetic dynamics of the system. The magnitude of harmonics can be increased or decreased by 60%, depending on the parameters of the MPS system. This work could provide insights on the design of magnetic bioassays, MPS and MPI systems where the magnitude of signals relies on the magnetic responses of SPIONs. Furthermore, it is shown that the clustering of SPIONs gives rise to a weaker magnetic response due to the presence of strong dipolar interactions.

2. Numerical Simulations

2.1 OOMMF Simulation Design

Object Oriented Micormagnetic Framework (OOMMF) was employed to simulate the demagnetization fields (dipolar fields) from two nearby SPIONs separated by distances of 20 nm, 80 nm, 120 nm, 200 nm and 300 nm, respectively. The building block of the simulation is the Landau-Lifshitz-Gilbert equation:

$$\frac{dM}{dt} = -|\gamma| M \times H_{eff} + \frac{\alpha}{M_s}\left(M \times \frac{dM}{dt}\right)$$

Where the saturation magnetization $M_s$ is $3.5 \times 10^5$ A/m (0.35 T), the Gilbert damping constant $\alpha$ is 0.1, and the gyromagnetic ratio $\gamma$ is $-2.21 \times 10^5$ mA$^{-1}$s$^{-1}$. Other material properties used in this simulation were the exchange stiffness constant A ($2.64 \times 10^{-11}$ J/m, $2.64 \times 10^{-6}$ erg/cm) and the crystalline anisotropy constant ($1.25 \times 10^4$ J/m$^3$, $1.25 \times 10^5$ ergs/cm$^3$). The cell size for the simulation was 2 nm.

2.2 Simulation Results

As shown in Figure 1 (a), when an external magnetic field is applied along the +x direction, the magnetizations of both SPIONs lie along the field direction. The demagnetization field at the center of the SPION is around $2.2 \times 10^5$ A/m. As the distance between the SPIONs increases, the overlap area between the demagnetization fields decreases (see Figures 1(b) - (f)). It was observed that the demagnetization fields from the SPIONs overlap at dipole-dipole distances below 120 nm, which indicates the strong dipolar interactions. When the spacing between SPIONs is larger than 200 nm, the demagnetization field from one SPION has negligible influence on the other, indicating much weaker or none dipolar interactions.

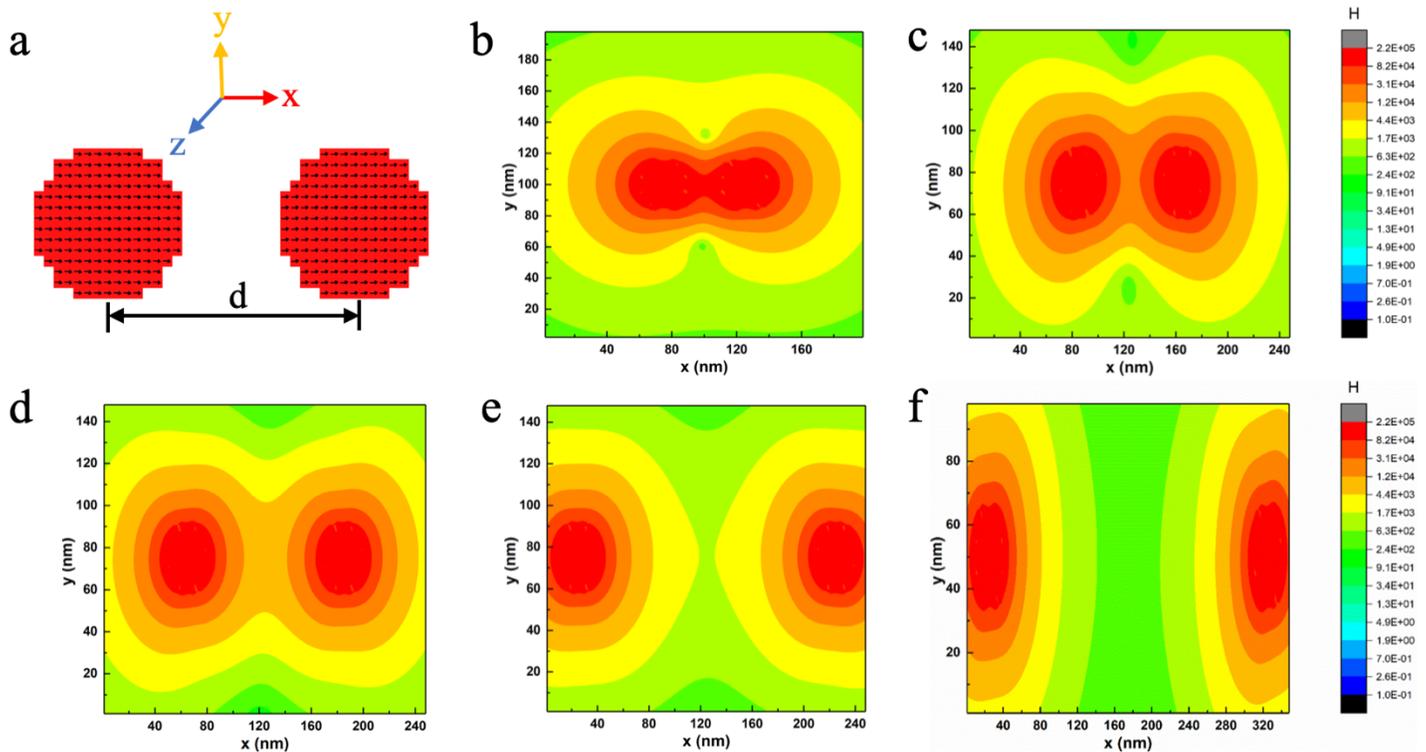

Figure 1. OOMMF simulation results on the dipolar interactions between two nearby SPIONs. (a) Two SPIONS with magnetizations lying along the external magnetic field direction (+x direction). (b)-(f) Demagnetization fields from two SPIONs that are separated by distances of 20 nm (b), 80 nm (c), 120 nm (d), 200 nm (e), and 300 nm (f). The demagnetization field unit is A/m in this figure.

3. Materials and Methods

3.1 Materials

In this work, we used superparamagnetic iron oxide nanoparticles (SPIONs) with an average magnetic core diameter of 30 nm (purchased from Ocean NanoTech, catalog#SHB-30). These single-core SPIONs are lipid coated iron oxide (γ-Fe$_2$O$_3$/Fe$_3$O$_4$) with biotin. The hydrodynamic size is about 20 nm larger than the magnetic core size measured by transmission electron microscopy (TEM, FEI Tecnai T12, 120 kV) and dynamic light scattering (DLS), see Supporting Information S1. The SHB-30 can be reconstituted in phosphate buffered saline (PBS).

3.2 Sample Preparation

SPION suspensions having nanoparticle volume fractions (VFs, ϕ) ranging from 0.001 to over 0.029 are experimentally studied in this paper. The SHB-30 specimen originally bought from Ocean NanoTech has a SPION concentration of 0.068 nmole/mL (volume fraction ϕ=0.000579), which is below the volume fraction range of interest in this work. At first, a 250 μL SHB-30 suspension was ultra-centrifuged at 11,000 RPM, acceleration 11,200 g for 45 min (PowerSpin$^{TM}$ BX Centrifuge). Then a 170 μL supernatant was removed, followed by 30-mins of water bath ultra-sonication (Branson 1510 Ultrasonic Cleaner) to make SPIONs evenly dispersed, resulting in the sample i as shown in Figure 2. Sample i contains 80 μL SPION suspension with a nanoparticle concentration of 0.2125 nmole/mL, volume fraction ϕ=0.0018 and an averaged inter-particle distance of 198 nm. MPS measurements were carried out on sample i. To get sample ii from sample i, ultra-centrifugation, supernatant removal, ultra-sonication processes were repeated (the condensation process (a) labeled in Figure 2). Sample ii was a 20 μL SPION suspension with a nanoparticle concentration of 0.85 nmole/mL, volume fraction ϕ=0.0072, and an averaged inter-particle distance of 125 nm. Then MPS measurements were carried out on sample ii. To get sample iii from sample ii, a dilution process (b) as labeled in Figure 2 was performed, which includes addition of a certain volume of PBS buffer to SPION suspension to dilute sample to a lower nanoparticle concentration (as well as a lower volume fraction ϕ and a larger averaged inter-particle distance), followed by 30-mins of water bath ultra-sonication. Samples iv - vii were prepared in the similar manner and each sample was measured by our lab-based MPS system. Details on the volume fractions and inter-particle distances of specimens i - vii can be found in Supporting Information S2.

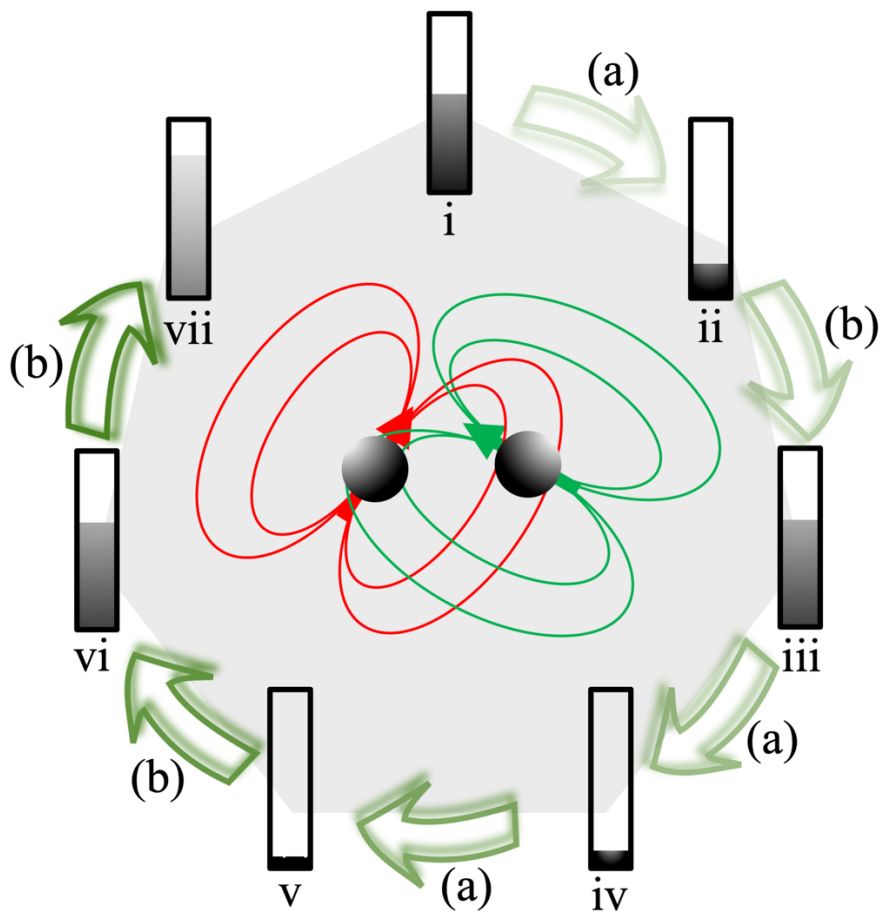

Figure 2. Samples: (i) 80 μL SPION suspension with volume fraction ϕ=0.0018; (ii) 20 μL SPION suspension with volume fraction ϕ=0.0072; (iii) 100 μL SPION suspension with volume fraction ϕ=0.0014; (iv) 5 μL SPION suspension with volume fraction ϕ=0.029; (v) Dried SPION powder with volume fraction ϕ>0.029; (vi) 120 μL SPION suspension with volume fraction ϕ=0.0012; (vii) 150 μL SPION suspension with volume fraction ϕ=0.001. Process (a) represents the condensation of SPION suspension: at first, the suspension is ultra-centrifuged at 11,000 RPM, acceleration 11,200 g for 45 min, then removal of a certain volume of supernatant, followed by 30-mins of water bath ultra-sonication to make SPIONs evenly dispersed. Process (b) represents the dilution of SPION suspension: initially, the suspension is added with a certain volume of PBS buffer to dilute sample to a lower nanoparticle concentration (as well as a lower volume fraction ϕ and a larger averaged inter-particle distance), followed by 30-mins of water bath ultra-sonication.

### 3.3 MPS Experimental Setups

In this work, a lab-based magnetic particle spectrometer (MPS) was used to monitor the magnetic responses of SPIONs with different particle concentrations c (varying from 0.1133 nmole/mL to 3.4 nmole/mL), volume fractions ϕ (varying from 0.001 to over 0.029), and inter-particle distances d (varying from below 78 nm to 245 nm). The MPS system setups have been reported by our previous works[31-36]. This system consists of: a PC with LabVIEW program to control the digital acquisition card (DAQ), carry out analog to digital convertor (ADC) and discrete Fourier Transform (DFT) on the analog signals that were sent back from the pick-up coils, two instrument amplifiers (IAs) receive commands from LabVIEW and send sinusoidal waves to drive coils, two sets of drive coils to generate high and low frequency alternating magnetic fields, one pair of differentially wound pick-up coils to collect the magnetic responses from SPIONs (Faraday's law of Induction), and a plastic vial to hold SPION sample. The schematic drawings of MPS system can be found in Supporting Information S3.

3.4 Experimental Methods

Two sinusoidal magnetic fields, one with high frequency $f_H$ (in this work we vary $f_H$ from 400 Hz to 20 kHz) but low amplitude $A_H = 17\ Oe$, the other with low frequency $f_L = 10\ Hz$ but high amplitude $A_L = 170\ Oe$ are applied to different SPION suspensions (samples i - vii)[35, 37-41]. Under the time-varying magnetic fields, the nonlinear magnetic responses of SPIONs induce electromotive force (EMF) in the pick-up coils (Faraday's Law of Induction), which is then sent back to DAQ and PC for harmonic signal extraction (details on the signal chain can be found in Supporting Information S3). The 3rd and the 5th harmonics at combinational frequencies $f_H \pm 2f_L$ and $f_H \pm 4f_L$ were used as indicators of the physical properties of SPION suspensions. For each sample, we carried out MPS measurements at different high frequencies $f_H$ (400 Hz to 20 kHz). At each frequency $f_H$, three consequent measurements were performed. In each measurement, the background noise floor was monitored for 10 s (10 data points). Then, the SPION sample in vial was inserted into the MPS system and the total signal was collected for another 10 s (10 data points). The voltage signals due to the nonlinear magnetic responses of SPIONs can be extracted by subtracting the background noise from the total signal according to the phasor theory[34-36] (details on the phasor theory can be found in Supporting Information S4).

4. MPS Measurement Results and Discussions

The magnetic responses of SPIONs are largely dependent on their magnetic core size, anisotropy parameter, hydrodynamic size[36, 42, 43], viscosity[32, 44] and temperature[45, 46] of suspension, as well as the effective magnetic field[2, 35] sensed by each SPION. In this work, we used the same batch of SPIONs, which could effectively minimize the variances in the magnetic core size, anisotropy parameter and hydrodynamic size from each measurement (see Supporting Information S5). Furthermore, the temperature and viscosity of these SPION suspensions were set identical so that the only factor that affects the magnetic responses would be the effective magnetic field sensed by each SPION (see Equations (6) & (13) from Supporting Information S6). The mathematical models of the magnetic responses of SPIONs and the induced voltage model from pick-up coils can be found in Supporting Information S6, and the mathematical models of the harmonic signals can be found in Supporting Information S7.

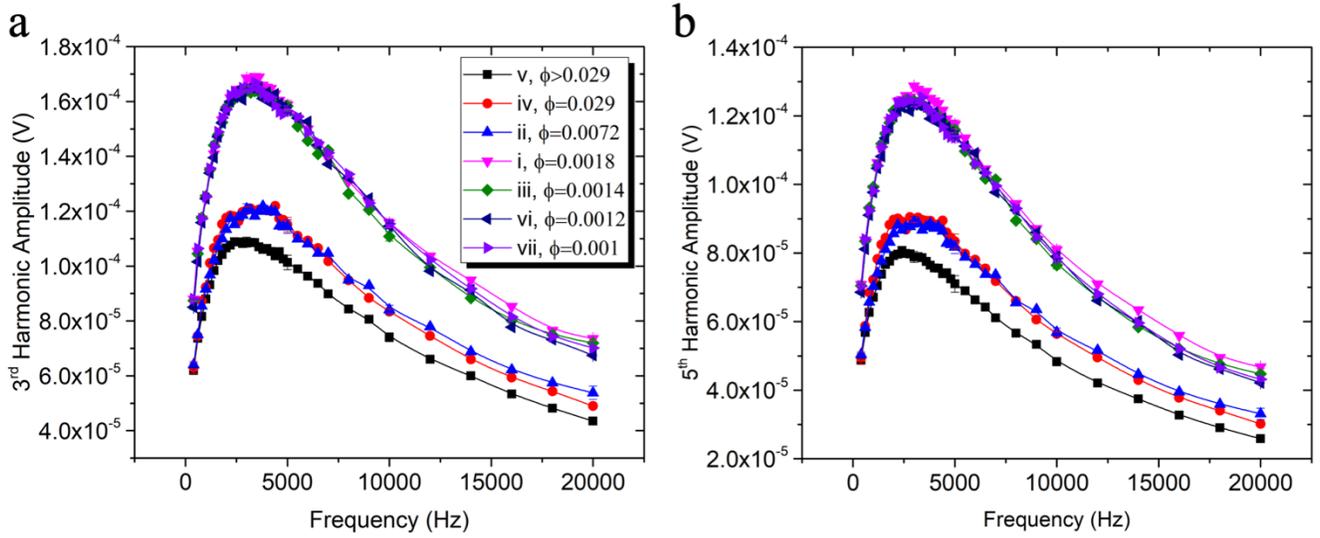

Figure 3. Measured (a) 3rd and (b) 5th harmonic amplitudes from samples i - vii as $f_H$ varies from 400 Hz to 20 kHz. Error bar represents the standard deviation. These seven SPION suspension samples come from the same batch and same amount of SPIONs but with different concentrations. Their magnetic responses can be divided into three groups: group 1, no dipolar interactions in samples i, iii, vi, and vii; group 2, moderate dipolar interactions in samples ii and iv; group 3, strong dipolar interactions in sample v.

As is shown in Figure 3, the averaged 3rd and 5th harmonic amplitudes are summarized as a function of the high frequencies $f_H$. By increasing SPION concentrations from 0.1133 nmole/mL (sample vii) to 3.4 nmole/mL (sample iv), the inter-particle distances decrease from 245 nm (sample vii) to 78 nm (sample iv). The dipolar magnetic fields (dipole-dipole interactions) generated by the nearby SPIONs increase with decreasing inter-particle distances, which, as a result, reduces the effective magnetic field on each SPION. A higher concentration of SPIONs in suspension leads to stronger dipolar interactions and lower harmonic amplitudes. In Figure 3, the magnetic responses of these SPION samples can be divided into 3 groups: 1) no dipolar interactions in samples i, iii, vi, and vii; 2) moderate dipolar interactions in samples ii and iv; 3) strong dipolar interactions in sample v. In group 1 (no dipolar interactions), the magnetic responses are identical from each SPION suspension even with the dilution of nanoparticles. In group 2 (with moderate dipolar interactions), the magnetic responses from sample ii (volume fraction ϕ=0.0072, averaged inter-particle distance d=125 nm) and sample iv (volume fraction ϕ=0.029, averaged inter-particle distance d=78 nm) are identical in the low $f_H$ range (below 7 kHz), and in the high $f_H$ range, sample iv shows smaller harmonic amplitudes due to the stronger dipolar interactions than sample ii. In group 3 (strong dipolar interactions), the densely stacked SPIONs dried powder leads to a stronger dipolar interaction than any other samples in this work. Consequently, weaker magnetic responses are detected in MPS system.

The harmonic signals from samples i to vii at four $f_H$ frequencies: 400 Hz, 3 kHz, 5 kHz, and 20 kHz are summarized in Figure 4. The group 1 samples (no dipolar interactions) are at similar amplitude levels at all frequencies. The group 2 and 3 samples (with moderate dipolar interactions and strong dipolar interactions) are at similar amplitude level in low frequency region (see Figure 4(a)). However, significant differences in the magnetic responses (i.e., the harmonic amplitudes) are found as we increase $f_H$ to 20 kHz (see Figure 4(c)). The effect of dipolar interaction reaches its maximum at $f_H = 3\ kHz$ when a harmonic magnitude difference of up to 60% is measured from the same batch of SPIONs with different volume fractions (different nanoparticle concentrations and inter-particle distances).

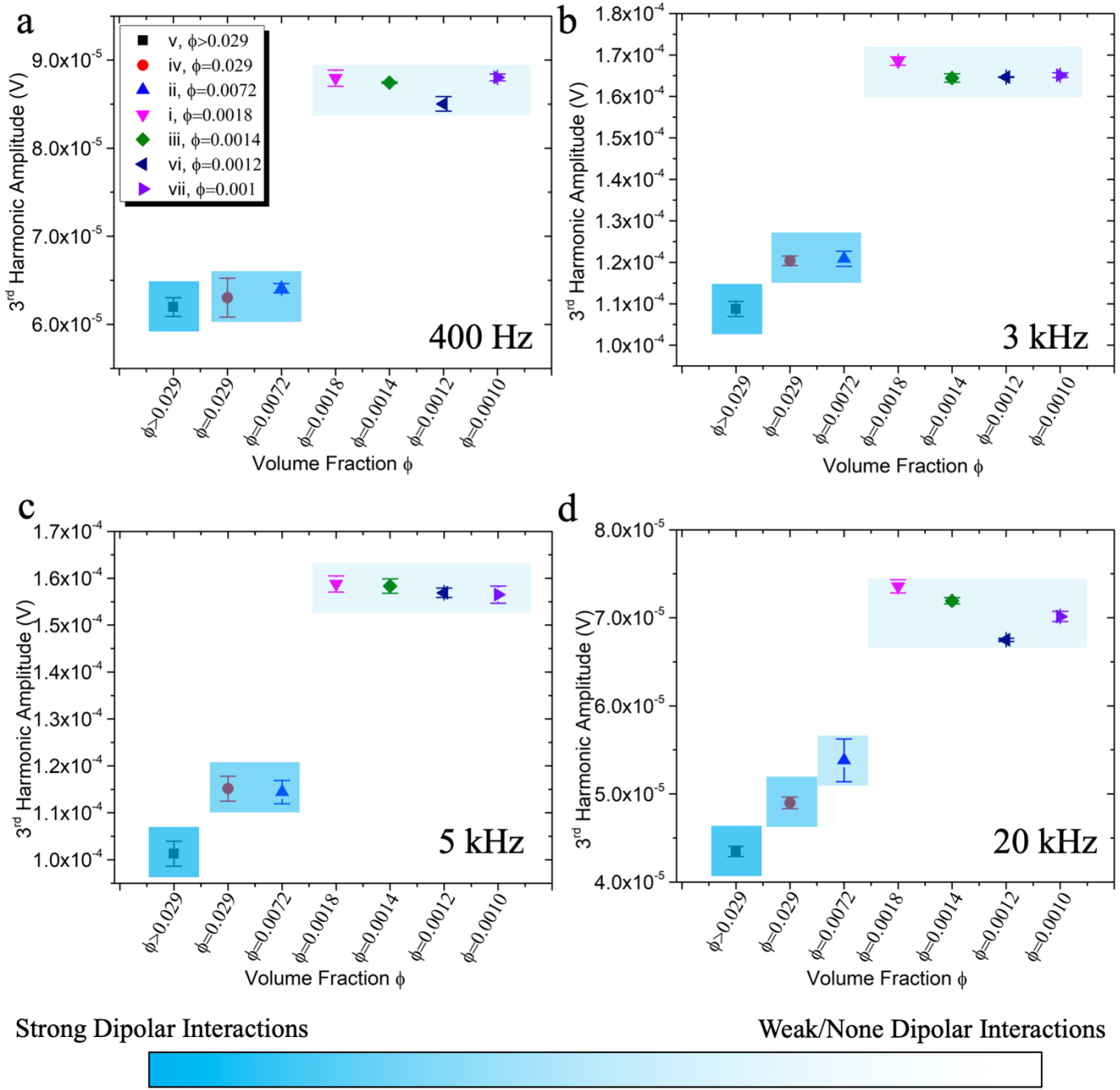

Figure 4. Measured 3rd harmonic amplitudes from samples i-vii at driving field frequencies of (a) $f_H = 400\ Hz$, (b) $f_H = 3\ kHz$, (c) $f_H = 5\ kHz$, and (d) $f_H = 20\ kHz$. Error bar represents the standard deviation.

5. Conclusions

In this paper, we reported a MPS-based method to experimentally investigate the effect of dipolar interactions on the magnetic dynamics of an ensemble of SPIONs. We measured the magnetic responses of seven SPION suspension samples that come from the same batch of nanoparticles with varying volume fractions $\phi$ from 0.001 to above 0.029 (inter-particle distances drop from 245 nm to below 78 nm, nanoparticle concentrations increase from 0.1133 to 3.4 nmole/mL). Our results show that even moderate changes in the SPION concentration could cause substantial effects on the magnetic dynamics of the SPION system, being capable of increasing or decreasing the harmonic signal magnitudes by 60%, depending on the values of MPS system parameters. Furthermore, our simulation results agree well with the MPS experimental results, where the strong dipolar interactions were observed at dipole-dipole distances below 120 nm whereas the dipolar field from one SPION has negligible influence on the other when the spacings between two SPIONs are larger than 200 nm, indicating much weaker or none dipolar interactions. Our OOMMF simulation results prove the feasibility of using the MPS system as a promising candidate for characterizing the dipolar interactions in any SPION systems.

These results give new insights on the design of magnetometers such as magnetic biosensors and MPI devices where the cumulative magnetic responses of an ensemble of SPIONs are of great importance in exerting magnetic signals for diagnosing and image reconstruction purposes. For example, in magnetometer-based bioassays (i.e., Superconducting Quantum Interference Devices (SQUID), Giant Magnetoresistive (GMR), etc.) and relaxometry-based bioassays (Magnetic Resonance Imaging (MRI), Nuclear Magnetic Resonance (NMR), etc.)[47-50], the assay results could largely be biased due to the fact that the magnetic characteristics of the SPION clusters differ from the non-interacting SPION system. In MPI where this non-invasive tomographic technique relies on the direct detection of SPION tracers' magnetic signals, however, the clustering of SPIONs at target tissues could lead to a loss of up to 60% in magnetic signals. On the other hand, the method reported in this work provides a possible way of using lab-based MPS for quantifying the degree of dipolar interactions in SPION systems as well as distinguishing the interacting and non-interacting SPION systems.

**Acknowledgements**

Portions of this work were conducted in the Minnesota Nano Center, which is supported by the National Science Foundation through the National Nano Coordinated Infrastructure Network (NNCI) under Award Number ECCS-1542202. Portions of this work were carried out in the Characterization Facility, University of Minnesota, a member of the NSF-funded Materials Research Facilities Network (www.mrfn.org) via the MRSEC program.

**Supporting Information**

Supporting Information S1. TEM and DLS characterization of SPIONs from SHB-30 specimen

Supporting Information S2. SPION volume fraction and the averaged inter-particle distances

Supporting Information S3. The lab-based MPS system and the signal chain

Supporting Information S4. The phasor theory

Supporting Information S5. Using same batch of SPIONs for MPS measurements

Supporting Information S6. Mathematical models of the magnetic responses of SPIONs and the induced voltage from pick-up coils

Supporting Information S7. Mathematical models of the harmonic signals from SPIONs

# AUTHOR INFORMATION

**Corresponding Author**

*E-mail: jpwang@umn.edu (J.-P. W)

**Notes**

The authors declare no competing financial interest.

**References**

1. Biehl, P.; von der Luhe, M.; Dutz, S.; Schacher, F. H., Synthesis, Characterization, and Applications of Magnetic Nanoparticles Featuring Polyzwitterionic Coatings. *Polymers* **2018,** *10* (1), 28.
2. Wu, K.; Wang, J.-P., Magnetic hyperthermia performance of magnetite nanoparticle assemblies under different driving fields. *AIP Advances* **2017,** *7* (5), 056327.
3. Coral, D. F.; Mendoza Zélis, P.; Marciello, M.; Morales, M. D. P.; Craievich, A.; Sanchez, F. H.; Fernández van Raap, M. B., On the effect of nanoclustering and dipolar interactions in heat generation for magnetic hyperthermia. *Langmuir* **2016**.
4. Coral, D. F.; Mendoza Zélis, P.; Marciello, M.; Morales, M. a. d. P.; Craievich, A.; Sánchez, F. H.; Fernández van Raap, M. B., Effect of nanoclustering and dipolar interactions in heat generation for magnetic hyperthermia. *Langmuir* **2016,** *32* (5), 1201-1213.
5. Hu, Y.; Ma, W.; Tao, M.; Zhang, X.; Wang, X.; Wang, X.; Chen, L., Decorated-magnetic-nanoparticle-supported bromine as a recyclable catalyst for the oxidation of sulfides. *Journal of Applied Polymer Science* **2018,** *135* (13).
6. Leshuk, T.; Holmes, A.; Ranatunga, D.; Chen, P. Z.; Jiang, Y.; Gu, F., Magnetic flocculation for nanoparticle separation and catalyst recycling. *Environmental Science: Nano* **2018**.
7. Moghaddam, F. M.; Saberi, V.; Kalhor, S.; Veisi, N., Palladium (II) Immobilized Onto the Glucose Functionalized Magnetic Nanoparticle as a New and Efficient Catalyst for the One-pot Synthesis of Benzoxazoles. *Applied Organometallic Chemistry* **2018**.
8. Zhou, Q.; Wan, Z.; Yuan, X.; Luo, J., A new magnetic nanoparticle-supported Schiff base complex of manganese: an efficient and recyclable catalyst for selective oxidation of alcohols. *Applied Organometallic Chemistry* **2016,** *30* (4), 215-220.

9. Yao, X.; Niu, X.; Ma, K.; Huang, P.; Grothe, J.; Kaskel, S.; Zhu, Y., Graphene Quantum Dots-Capped Magnetic Mesoporous Silica Nanoparticles as a Multifunctional Platform for Controlled Drug Delivery, Magnetic Hyperthermia, and Photothermal Therapy. *Small* **2017,** *13* (2).

10. Chowdhuri, A. R.; Bhattacharya, D.; Sahu, S. K., Magnetic nanoscale metal organic frameworks for potential targeted anticancer drug delivery, imaging and as an MRI contrast agent. *Dalton Transactions* **2016,** *45* (7), 2963-2973.

11. Hervault, A.; Dunn, A. E.; Lim, M.; Boyer, C.; Mott, D.; Maenosono, S.; Thanh, N. T., Doxorubicin loaded dual pH-and thermo-responsive magnetic nanocarrier for combined magnetic hyperthermia and targeted controlled drug delivery applications. *Nanoscale* **2016,** *8* (24), 12152-12161.

12. Wang, G.; Ma, Y.; Wei, Z.; Qi, M., Development of multifunctional cobalt ferrite/graphene oxide nanocomposites for magnetic resonance imaging and controlled drug delivery. *Chemical Engineering Journal* **2016,** *289*, 150-160.

13. Huang, J.; Li, Y.; Orza, A.; Lu, Q.; Guo, P.; Wang, L.; Yang, L.; Mao, H., Magnetic Nanoparticle Facilitated Drug Delivery for Cancer Therapy with Targeted and Image-Guided Approaches. *Advanced functional materials* **2016,** *26* (22), 3818-3836.

14. Estelrich, J.; Escribano, E.; Queralt, J.; Busquets, M. A., Iron oxide nanoparticles for magnetically-guided and magnetically-responsive drug delivery. *International journal of molecular sciences* **2015,** *16* (4), 8070-8101.

15. Krishna, V. D.; Wu, K.; Perez, A. M.; Wang, J. P., Giant Magnetoresistance-based Biosensor for Detection of Influenza A Virus. *Frontiers in Microbiology* **2016,** *7*, 8.

16. Wang, Y.; Wang, W.; Yu, L. N.; Tu, L.; Feng, Y. L.; Klein, T.; Wang, J. P., Giant magnetoresistive-based biosensing probe station system for multiplex protein assays. *Biosensors & Bioelectronics* **2015,** *70*, 61-68.

17. Rizzi, G.; Lee, J. R.; Dahl, C.; Guldberg, P.; Dufva, M.; Wang, S. X.; Hansen, M. F., Simultaneous Profiling of DNA Mutation and Methylation by Melting Analysis Using Magnetoresistive Biosensor Array. *Acs Nano* **2017,** *11* (9), 8864-8870.

18. Choi, J.; Gani, A. W.; Bechstein, D. J.; Lee, J.-R.; Utz, P. J.; Wang, S. X., Portable, one-step, and rapid GMR biosensor platform with smartphone interface. *Biosensors and Bioelectronics* **2016,** *85*, 1-7.

19. Wells, J.; Paysen, H.; Kosch, O.; Trahms, L.; Wiekhorst, F., Temperature dependence in magnetic particle imaging. *AIP Advances* **2018,** *8* (5), 056703.

20. Zanganeh, S.; Aieneravaie, M.; Erfanzadeh, M.; Ho, J.; Spitler, R., Magnetic Particle Imaging (MPI). In *Iron Oxide Nanoparticles for Biomedical Applications*, Elsevier: 2018; pp 115-133.

21. Ludewig, P.; Gdaniec, N.; Sedlacik, J.; Forkert, N. D.; Szwargulski, P.; Graeser, M.; Adam, G.; Kaul, M. G.; Krishnan, K. M.; Ferguson, R. M. J. A. n., Magnetic particle imaging for real-time perfusion imaging in acute stroke. **2017,** *11* (10), 10480-10488.

22. Qiao, Y.; Gumin, J.; MacLellan, C. J.; Gao, F.; Bouchard, R.; Lang, F. F.; Stafford, R. J.; Melancon, M. P., Magnetic resonance and photoacoustic imaging of brain tumor mediated by mesenchymal stem cell labeled with multifunctional nanoparticle introduced via carotid artery injection. *Nanotechnology* **2018,** *29* (16), 165101.

23. Ravanshad, R.; Zadeh, A. K.; Amani, A. M.; Mousavi, S. M.; Hashemi, S. A.; Dashtaki, A. S.; Mirzaei, E.; Zare, B., Application of nanoparticles in cancer detection by Raman scattering based techniques. *Nano Reviews & Experiments* **2017,** *9*.

24. Shen, Z.; Wu, A.; Chen, X., Iron oxide nanoparticle based contrast agents for magnetic resonance imaging. *Molecular pharmaceutics* **2016,** *14* (5), 1352-1364.

25. Branquinho, L. C.; Carrião, M. S.; Costa, A. S.; Zufelato, N.; Sousa, M. H.; Miotto, R.; Ivkov, R.; Bakuzis, A. F., Effect of magnetic dipolar interactions on nanoparticle heating efficiency: Implications for cancer hyperthermia. *Scientific reports* **2013,** *3*, 2887.

26. Kechrakos, D.; Trohidou, K., Competition between dipolar and exchange interparticle interactions in magnetic nanoparticle films. *Journal of magnetism and magnetic materials* **2003,** *262* (1), 107-110.

27. Landi, G. T., Role of dipolar interaction in magnetic hyperthermia. *Physical Review B* **2014,** *89* (1), 014403.

28. García-Otero, J.; Porto, M.; Rivas, J.; Bunde, A., Influence of dipolar interaction on magnetic properties of ultrafine ferromagnetic particles. *Physical review letters* **2000,** *84* (1), 167.

29. Donahue, M. J. *OOMMF user's guide, version 1.0*; 1999.

30. Donahue, M.; Porter, D.; Lau, J.; McMichael, R., in Interagency Report NISTIR 6376 (National Institute of Standards and Technology, Gaithersburg, MD). *NIST J. Res.* **114**, 57-67.

31. Wu, K.; Yu, L.; Zheng, X.; Wang, Y.; Feng, Y.; Tu, L.; Wang, J.-P. In *Viscosity effect on the brownian relaxation based detection for immunoassay applications*, 2014 36th Annual International Conference of the IEEE Engineering in Medicine and Biology Society, IEEE: 2014; pp 2769-2772.

32. Wu, K.; Liu, J.; Wang, Y.; Ye, C.; Feng, Y.; Wang, J.-P., Superparamagnetic nanoparticle-based viscosity test. *Applied Physics Letters* **2015,** *107* (5), 053701.

33. Wu, K.; Batra, A.; Jain, S.; Wang, J.-P., Magnetization Response Spectroscopy of Superparamagnetic Nanoparticles Under Mixing Frequency Fields. *IEEE Transactions on magnetics* **2016,** *52* (7).

34. Tu, L.; Wu, K.; Klein, T.; Wang, J.-P., Magnetic nanoparticles colourization by a mixing-frequency method. *Journal of Physics D: Applied Physics* **2014,** *47* (15), 155001.

35. Wu, K.; Tu, L.; Su, D.; Wang, J.-P., Magnetic dynamics of ferrofluids: mathematical models and experimental investigations. *Journal of Physics D: Applied Physics* **2017,** *50* (8), 085005.

36. Wu, K.; Schliep, K.; Zhang, X.; Liu, J.; Ma, B.; Wang, J. P., Characterizing Physical Properties of Superparamagnetic Nanoparticles in Liquid Phase Using Brownian Relaxation. *Small* **2017,** *13* (22).

37. Krause, H.-J.; Wolters, N.; Zhang, Y.; Offenhäusser, A.; Miethe, P.; Meyer, M. H.; Hartmann, M.; Keusgen, M., Magnetic particle detection by frequency mixing for immunoassay applications. *Journal of magnetism and magnetic materials* **2007,** *311* (1), 436-444.

38. Nikitin, P. I.; Vetoshko, P. M.; Ksenevich, T. I., New type of biosensor based on magnetic nanoparticle detection. *Journal of Magnetism and Magnetic Materials* **2007,** *311* (1), 445-449.

39. Tu, L.; Jing, Y.; Li, Y.; Wang, J.-P., Real-time measurement of Brownian relaxation of magnetic nanoparticles by a mixing-frequency method. *Applied Physics Letters* **2011,** *98* (21), 213702.

40. Tu, L.; Feng, Y.; Klein, T.; Wang, W.; Wang, J.-P., Measurement of Brownian Relaxation of Magnetic Nanoparticle by a Multi-Tone Mixing-Frequency Method. *Magnetics, IEEE Transactions on* **2012,** *48* (11), 3513-3516.

41. Tu, L.; Klein, T.; Wang, W.; Feng, Y.; Wang, Y.; Wang, J.-P., Measurement of Brownian and Néel relaxation of magnetic nanoparticles by a mixing-frequency method. *Magnetics, IEEE Transactions on* **2013,** *49* (1), 227-230.

42. Zhang, X.; Reeves, D. B.; Perreard, I. M.; Kett, W. C.; Griswold, K. E.; Gimi, B.; Weaver, J. B., Molecular sensing with magnetic nanoparticles using magnetic spectroscopy of nanoparticle Brownian motion. *Biosensors and Bioelectronics* **2013,** *50*, 441-446.

43. Rauwerdink, A. M.; Weaver, J. B., Measurement of molecular binding using the Brownian motion of magnetic nanoparticle probes. *Applied Physics Letters* **2010,** *96* (3), 033702.

44. Weaver, J. B.; Harding, M.; Rauwerdink, A. M.; Hansen, E. W. In *The effect of viscosity on the phase of the nanoparticle magnetization induced by a harmonic applied field*, SPIE Medical Imaging, International Society for Optics and Photonics: 2010; pp 762627-762627-8.

45. Rauwerdink, A. M.; Hansen, E. W.; Weaver, J. B., Nanoparticle temperature estimation in combined ac and dc magnetic fields. *Physics in medicine and biology* **2009,** *54* (19), L51.

46. Weaver, J. B.; Rauwerdink, A. M.; Hansen, E. W., Magnetic nanoparticle temperature estimation. *Medical physics* **2009,** *36* (5), 1822-1829.

47. Eberbeck, D.; Wiekhorst, F.; Steinhoff, U.; Trahms, L., Aggregation behaviour of magnetic nanoparticle suspensions investigated by magnetorelaxometry. *Journal of Physics: Condensed Matter* **2006,** *18* (38), S2829.

48. Eberbeck, D.; Trahms, L., Experimental investigation of dipolar interaction in suspensions of magnetic nanoparticles. *Journal of Magnetism and Magnetic Materials* **2011,** *323* (10), 1228-1232.

49. Spasova, M.; Wiedwald, U.; Ramchal, R.; Farle, M.; Hilgendorff, M.; Giersig, M., Magnetic properties of arrays of interacting Co nanocrystals. *Journal of magnetism and magnetic materials* **2002,** *240* (1-3), 40-43.

50. Yang, Z.; Shi, F.; Wang, P.; Raatz, N.; Li, R.; Qin, X.; Meijer, J.; Duan, C.; Ju, C.; Kong, X., Detection of magnetic dipolar coupling of water molecules at the nanoscale using quantum magnetometry. *Physical Review B* **2018,** *97* (20), 205438.

# Supporting Information

# Investigating the Effect of Magnetic Dipole-Dipole Interaction on Magnetic Particle Spectroscopy (MPS): Implications for Magnetic Nanoparticle-based Bioassays and Magnetic Particle Imaging (MPI)

Kai Wu[†], Diqing Su[‡], Renata Saha[†], Jinming Liu[†], and Jian-Ping Wang[†,*]

[†]Department of Electrical and Computer Engineering, University of Minnesota, Minneapolis, Minnesota 55455, USA

[‡]Department of Chemical Engineering and Material Science, University of Minnesota, Minneapolis, Minnesota 55455, USA

*Corresponding author E-mail: jpwang@umn.edu

**Supporting Information S1. TEM and DLS characterization of SPIONs from SHB-30 specimen.**

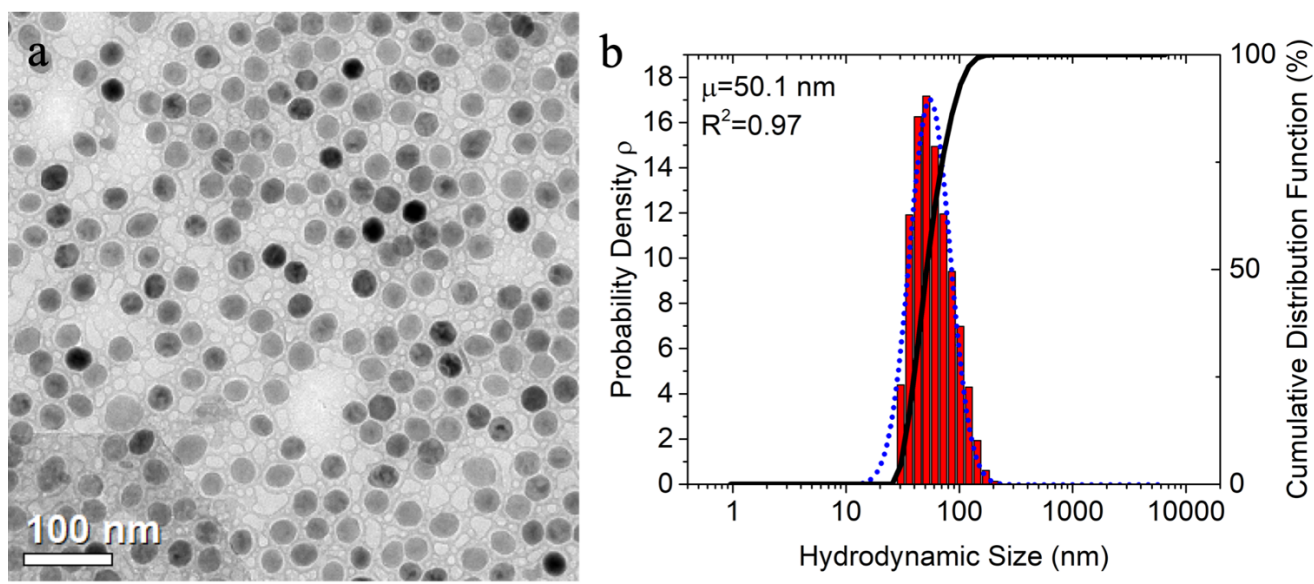

Figure S1. (a) TEM image of the SPIONs from SHB-30 specimen. (b) Statistical hydrodynamic size distribution collected using DLS. Solid black line is cumulative distribution curve and dash blue line is the lognormal curve fitting. SPIONs show an average hydrodynamic size of 50.1 nm.

**Supporting Information S2. SPION volume fraction and the averaged inter-particle distances.**

Table S1. Nanoparticle volume fractions and the averaged inter-particle distances

| Sample Label* | Volume | SPION Concentration c (nmole/mL) | Volume Fraction ($\phi$)** | Averaged Inter-particle Distance d (nm) |
|---|---|---|---|---|
| i | 80 µL | 0.2125 | 0.0018 | 198 |
| ii | 20 µL | 0.85 | 0.0072 | 125 |
| iii | 100 µL | 0.17 | 0.0014 | 214 |
| iv | 5 µL | 3.4 | 0.029 | 78 |
| v | Dried powder | NA | > 0.029 | < 78 |
| vi | 120 µL | 0.1417 | 0.0012 | 227 |
| vii | 150 µL | 0.1133 | 0.001 | 245 |

*Sample labels correspond to the marks in Figure 2 from manuscript.

**Volume fraction (VF) is calculated by $\phi = \frac{\Sigma_k V_k}{V_s}$, where $V_k$ is the magnetic core volume of SPION $k$ and $V_s$ is the volume of suspension.

**Supporting Information S3. The lab-based MPS system and the signal chain.**

As shown in Figure S2, the MPS system consists of a PC with LabVIEW program to control the generation of two sinusoidal waves that are sent to DAQ and instrumental amplifiers (IAs), analog to digital convertor (ADC) and discrete Fourier Transform (DFT) to analyze the collected signal from pick-up coils. Two sinusoidal signals are sent to drive coils to generate two sinusoidal magnetic fields, one with high frequency but low amplitude $A_H \sin(2\pi f_H t + \varphi_H)$ and the other with low frequency but high amplitude $A_L \sin(2\pi f_L t + \varphi_L)$, where $A_H$ and $A_L$ are the amplitudes of high and low frequency magnetic fields, $f_H$ and $f_L$ are the frequencies of high and low frequency magnetic fields, $\varphi_H$ and $\varphi_L$ are the phases of high and low frequency magnetic fields, and $t$ is time parameter. One pair of differentially wound pick-up coils is put in the center of two outer coils, and the plastic vial with SPION suspension inside will be inserted into the pick-up coils. The specially designed plastic vial can be filled up to 300 µL liquid. As the time-varying sinusoidal magnetic fields applied to the SPIONs, the nonlinear magnetic responses of SPIONs generate time-varying magnetic fields that can be sensed by the pick-up coils according to the Faraday's law of Induction. The real-time voltage signal from pick-up coils is sent back to DAQ and LabVIEW after a bandpass filter (BPF). The harmonic signals at combinational frequencies $f_H \pm 2f_L$ (the 3rd harmonics) and $f_H \pm 4f_L$ (the 5th harmonics) are extracted after DFT for analysis.

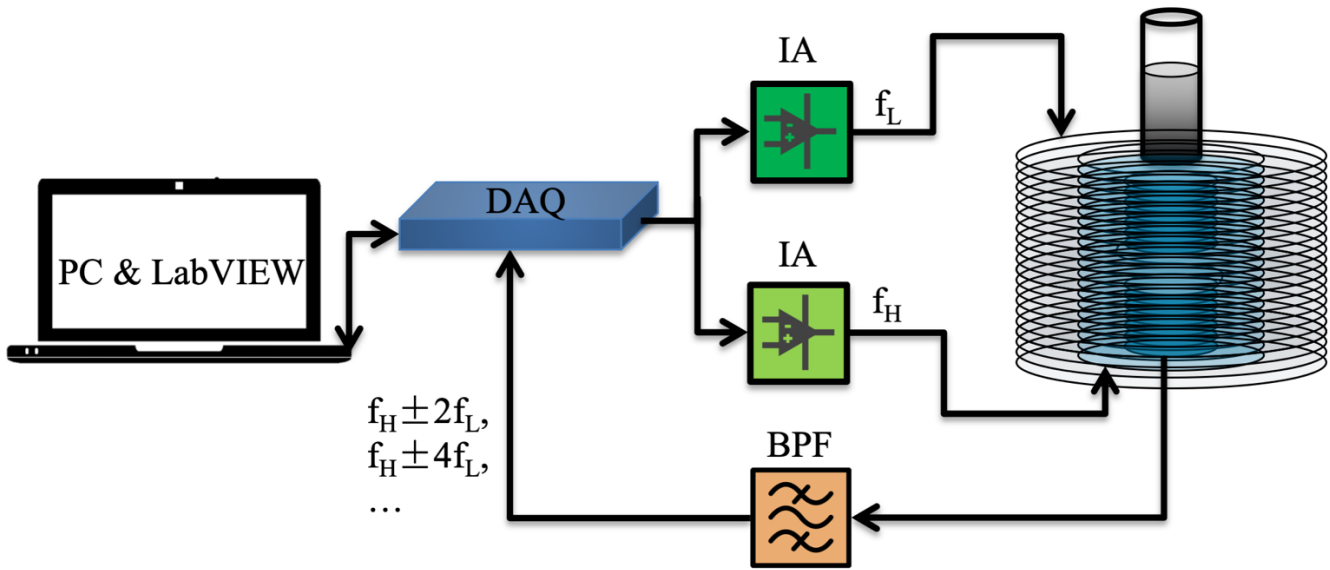

Figure S2. Schematic view of MPS system.

**Supporting Information S4. The phasor theory.**

The voltage caused by SPIONs at a certain frequency is represented by a phasor: $A \cdot e^{j(\omega t+\varphi)}$ (or expressed as $A\angle\varphi$), where $\omega$ is the angular frequency of driving field, $\varphi$ is the phase angle, and $j = \sqrt{-1}$.

In our experimental setup, two sinusoidal magnetic fields are applied by passing alternating currents (ACs) to the driving coils. Firstly, the background noise is collected with the external magnetic fields on. The background noise can be expressed as $A_{Noise}e^{j\varphi_{Noise}}$. Secondly, a plastic vial containing SPION sample is inserted into the MPS system and the total signal is collected. The total signal is expressed as $A_{TOT}e^{j\varphi_{TOT}}$. This total signal is the sum of two phasors: the background noise and the SPIONs (namely, $A_{MNP}e^{j\varphi_{MNP}}$).

So,
$$A_{Noise}e^{j\varphi_{Noise}} + A_{MNP}e^{j\varphi_{MNP}} = A_{TOT}e^{j\varphi_{TOT}}, (1)$$

which reduces to an equation set:
$$\begin{cases} A_{Noise} \times \cos\varphi_{Noise} + A_{MNP} \times \cos\varphi_{MNP} = A_{TOT} \times \cos\varphi_{TOT} \\ A_{Noise} \times \sin\varphi_{Noise} + A_{MNP} \times \sin\varphi_{MNP} = A_{TOT} \times \sin\varphi_{TOT} \end{cases} (2)$$

By solving the equation set above, we can get the harmonic amplitude $A_{MNP}$ and phase $\varphi_{MNP}$ of SPIONs at different frequencies.

**Supporting Information S5. Using same batch of SPIONs for MPS measurements.**

To minimize the variances in the magnetic core size, anisotropy parameter, and hydrodynamic size from different batches of SPIONs, the same batch of SPIONs are used for all the MPS measurements by repeating the dilution and condensation processes. The cycling steps can be found in Figure 2 from manuscript. We started with sample i (80 μL in volume, with nanoparticle concentration c=0.2125 nmole/mL, volume fraction ϕ=0.0018, and averaged inter-particle distance d=198 nm), and sample iii is achieved after one condensation process and one dilution process on sample i. Sample iii (100 μL in volume, with nanoparticle concentration c=0.17 nmole/mL, volume fraction ϕ=0.0014, and averaged inter-particle distance d=214 nm) consists of the same batch of SPIONs from sample i and shows the similar magnetic responses with no dipolar interactions. After several condensation and dilutions process, we can get the similar magnetic responses from SPIONs under the conditions of no dipolar interactions (samples i, iii, vi, and vii) or with dipolar interactions (samples ii, iv, and v) as shown in Figure S3.

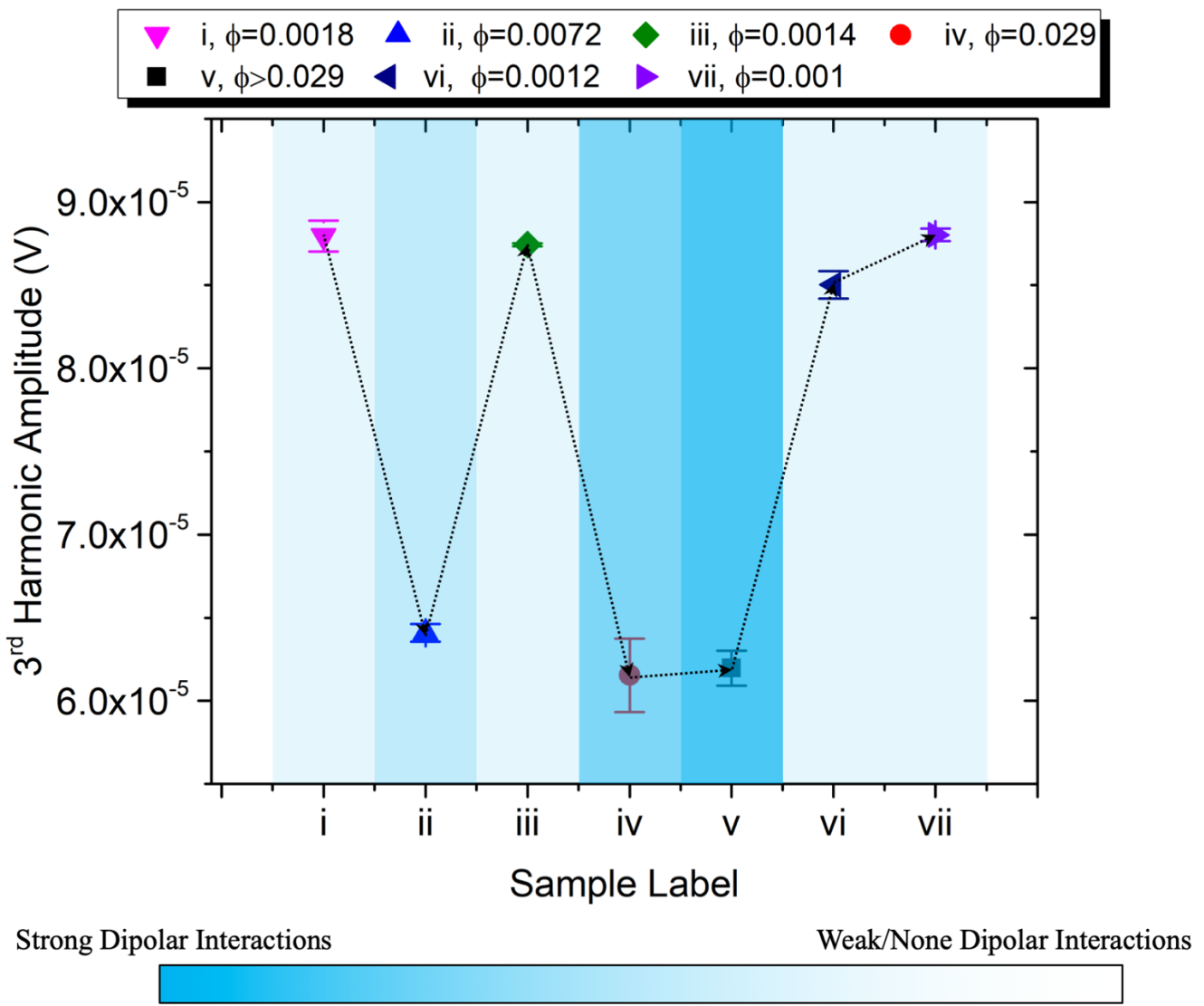

Figure S3. The 3rd harmonic amplitudes from SPION samples i – vii measured at $f_H = 400\ Hz$.

## Supporting Information S6. Mathematical models of the magnetic responses of SPIONs and the induced voltage from pick-up coils.

In the presence of sinusoidal magnetic fields, SPIONs are magnetized and their magnetic moments tend to align with the external magnetic fields.

For a SPION suspension of monodispersed, noninteracting SPIONs, the magnetic response can be expressed as the Langevin function:

$$M_D(t) = \mu c L(\xi), (3)$$

where,

$$L(\xi) = \coth \xi - \frac{1}{\xi} (4)$$

$$\xi = \frac{\mu H(t)}{k_B T} (5)$$

The SPIONs are characterized by magnetic core diameter $D$, saturation magnetization $M_s$ and concentration $c$. Assume SPIONs are spherical **without interparticle interactions (i.e., dipolar interactions)**. The magnetic moment of each nanoparticle is $\mu = M_s V_c = M_s \pi D^3/6$, $V_c$ is volume of the magnetic core, $\xi$ is the ratio of magnetic energy over thermal energy, $k_B$ is Boltzmann constant, and $T$ is the absolute temperature in Kelvin. $H(t) = A_H \sin(2\pi f_H t + \varphi_H) + A_L \sin(2\pi f_L t + \varphi_L)$ are the external magnetic fields.

On the other hand, for the densely distributed SPIONs **with interparticle interactions**, the effective magnetic field applied to each SPION is the sum of externally applied magnetic fields $\vec{H}(t)$ and the cumulative dipolar magnetic fields, $\vec{H}(t)|_{dipole}$, also called stray field or demagnetizing field, that exerted by nearby SPIONs, expressed as:

$$\vec{H}(t)|_{eff}^i = \vec{H}(t) + \vec{H}(t)|_{dipole} = [A_H \sin(2\pi f_H t + \varphi_H) + A_L \sin(2\pi f_L t + \varphi_L)]\hat{z} + \sum_j \vec{H}(t)|_{dipol}^j (6)$$

$$\vec{H}(t)|_{dipol}^j (\vec{r}_{i,j}) = \frac{1}{4\pi} \left( \frac{3\vec{r}_{i,j} \cdot (\vec{\mu}_j \cdot \vec{r}_{i,j})}{|\vec{r}_{i,j}|^5} - \frac{\vec{\mu}_j}{|\vec{r}_{i,j}|^3} \right) (7)$$

Where $H(t)|_{eff}^i$ is the effective magnetic field on the $i^{th}$ SPION, $\sum_j \vec{H}(t)|_{dipol}^j$ is the cumulative dipolar magnetic field generated by the nearby SPIONs, $\vec{H}(t)|_{dipol}^j (\vec{r}_{i,j})$ is the dipolar field of the $j^{th}$ SPION on the $i^{th}$ SPION, $\vec{r}_{i,j}$ is the length of the vector connecting SPIONs $i$ and $j$, and $\vec{\mu}_j$ is the magnetic moment vector of the $j^{th}$ SPION, $\hat{z}$ as the unit vector along the externally applied magnetic field.

The energy of SPION $i$ consists of the anisotropy energy, Zeeman energy, and dipolar interaction energy [1]. By defining $\hat{\mu}_i$ as the unit vector along the direction of magnetic moment $\vec{\mu}_i$ of SPION $i$, $\hat{n}_i$ as the unit vector along the easy axis, $\hat{z}$ as the unit vector along the externally applied magnetic field, $\vec{r}_{i,j}$ as vector connecting magnetic moments $\hat{\mu}_i$ and $\hat{\mu}_j$ [2], $\vec{H}(t)|_{eff}^i$ as the sum of external magnetic field and dipolar fields from nearby particles sensed by particle $i$.

Anisotropy energy $E_A^{(i)}$ is:

$$E_A^{(i)} = -K_{eff}V_c(\hat{u}_i \cdot \hat{n}_i)^2 \quad (8)$$

Zeeman energy $E_Z^{(i)}$ is:

$$E_Z^{(i)} = -\vec{\mu}_i \cdot \mu_0 \vec{H}(t) \quad (9)$$

Dipolar interaction energy $E_D^{(i,j)}$ is:

$$E_D^{(i,j)} = -\vec{\mu}_i \cdot \mu_0 \vec{H}(t)\Big|_{dipol}^j (\vec{r}_{i,j}) = \frac{\mu_0}{4\pi}\left(\frac{\vec{\mu}_i \cdot \vec{\mu}_j}{|\vec{r}_{i,j}|^3} - \frac{3(\vec{\mu}_i \cdot \vec{r}_{i,j})(\vec{\mu}_j \cdot \vec{r}_{i,j})}{|\vec{r}_{i,j}|^5}\right) \quad (10)$$

By taking dipolar interaction into consideration, the total energy of particle $i$ is:

$$E^{(i)} = E_A^{(i)} + E_Z^{(i)} + E_D^{(i,j)} \quad (11)$$

Rewrite above equation by adding up equivalent dipolar field and external magnetic field[2]:

$$E^{(i)} = -K_{eff}V_c(\hat{u}_i \cdot \hat{n}_i)^2 - \vec{\mu}_i \cdot \mu_0 \vec{H}(t)\Big|_{eff}^i \quad (12)$$

where

$$\mu_0 \vec{H}(t)\Big|_{eff}^i = \mu_0 \vec{H}(t) - \frac{\mu_0}{4\pi}\sum_{j \neq i}\left(\frac{\vec{\mu}_j}{|\vec{r}_{i,j}|^3} - \frac{3\vec{r}_{i,j}(\vec{\mu}_j \cdot \vec{r}_{i,j})}{|\vec{r}_{i,j}|^5}\right) \quad (13)$$

Hence, the dipole-dipole interaction and external magnetic field have changed the initially reported magnetic anisotropy barrier, as shown in Figure S4.

In this work, we have monitored the magnetic responses of SPION samples i - vii, where the MPS results from this work indicate that: samples ii, iv, and v are in status (c) from Figure S4 (with dipolar interactions), samples i, iii, vi, and vii are in status (b) from Figure S4 (without dipolar interactions).

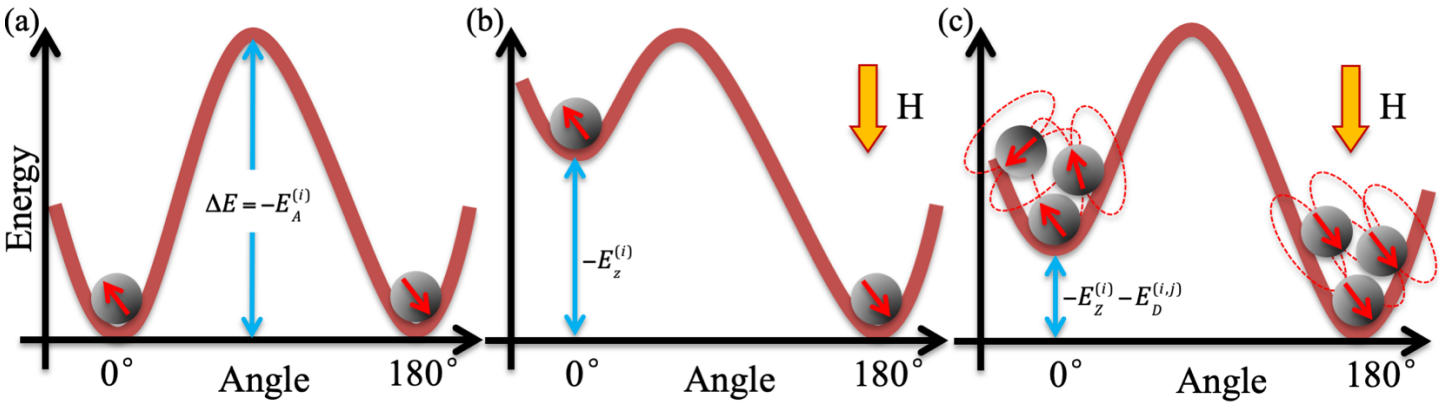

Figure S4. Schematic view of energy barriers (a) without external magnetic fields and dipolar interactions; (b) with external magnetic fields only; (c) with both external magnetic fields and dipolar interactions.

**Supporting Information S7. Mathematical models of the harmonic signals from SPIONs.**

Taylor expansion of the $\vec{M}_D(t)$ shows the major mixing frequency components (harmonic signals) due to the nonlinear magnetic responses of SPIONs:

$$\frac{\vec{M}_D(t)}{\mu c} = L\left(\frac{\mu \vec{H}(t)}{k_B T}\right)$$

$$= \frac{1}{3}\left(\frac{\mu}{k_B T}\right)\vec{H}(t) - \frac{1}{45}\left(\frac{\mu}{k_B T}\right)^3 \vec{H}(t)^3 + \frac{2}{945}\left(\frac{\mu}{k_B T}\right)^5 \vec{H}(t)^5 + \cdots$$

$$= \cdots + \left[-\frac{1}{60} A_H A_L^2 \left(\frac{\mu}{k_B T}\right)^3 + o\left(\left(\frac{\mu}{k_B T}\right)^3\right)\right] \times \cos[2\pi(f_H \pm 2f_L)t]\hat{z}$$

$$+ \left[\frac{1}{1512} A_H A_L^4 \left(\frac{\mu}{k_B T}\right)^5 + o\left(\left(\frac{\mu}{k_B T}\right)^5\right)\right] \times \cos[2\pi(f_H \pm 4f_L)t]\hat{z}$$

$$+ \cdots$$

(14)

For an assembly of SPIONs with diameter $D$, suspension volume of $V$, and nanoparticle concentration of $c$, confined in a cylindrical vial inside a pair of pick-up coils (see Figure S2), the overall magnetic moment from the SPION sample consisting the 3rd and the 5th harmonic components is expressed as (in *A/m*):

$$\vec{M}_D(t)\big|_{3rd} \approx \mu c V \left[-\frac{1}{60} A_H A_L^2 \left(\frac{\mu}{k_B T}\right)^3 + o\left(\left(\frac{\mu}{k_B T}\right)^3\right)\right] \times \cos[2\pi(f_H + 2f_L)t]\hat{z} \quad (15)$$

$$\vec{M}_D(t)\big|_{5th} \approx \mu c V \left[\frac{1}{1512} A_H A_L^4 \left(\frac{\mu}{k_B T}\right)^5 + o\left(\left(\frac{\mu}{k_B T}\right)^5\right)\right] \times \cos[2\pi(f_H + 4f_L)t]\hat{z} \quad (16)$$

According to Faraday's law of Induction, the induced time-varying 3rd and 5th harmonic voltages sensed by the pick-up coils are (in *Volt*):

$$u(t)|_{3rd} = -NS \frac{d\left(\mu_0 \vec{M}_D(t)\big|_{3rd}\right)}{dt} \quad (17)$$

$$u(t)|_{5th} = -NS \frac{d\left(\mu_0 \vec{M}_D(t)\big|_{5th}\right)}{dt} \quad (18)$$

where $N$ is the number of winding turns in pick-up coils, $S$ is the area contained by the pick-up coils [3-5].

By replacing $\vec{H}(t)$ with the effective magnetic field $\vec{H}(t)\big|_{eff}^i$ (which considers the dipolar fields), the time-varying 3rd and 5th harmonic voltages will be altered, as a result, the harmonic amplitudes after DFT will be altered.

**References**

1. García-Otero, J.; Porto, M.; Rivas, J.; Bunde, A., Influence of dipolar interaction on magnetic properties of ultrafine ferromagnetic particles. *Physical review letters* **2000,** *84* (1), 167.
2. Andersson, J.-O.; Djurberg, C.; Jonsson, T.; Svedlindh, P.; Nordblad, P., Monte Carlo studies of the dynamics of an interacting monodispersive magnetic-particle system. *Physical Review B* **1997,** *56* (21), 13983.
3. Wu, K.; Schliep, K.; Zhang, X.; Liu, J.; Ma, B.; Wang, J. P., Characterizing Physical Properties of Superparamagnetic Nanoparticles in Liquid Phase Using Brownian Relaxation. *Small* **2017,** *13* (22).
4. Wu, K.; Tu, L.; Su, D.; Wang, J.-P., Magnetic dynamics of ferrofluids: mathematical models and experimental investigations. *Journal of Physics D: Applied Physics* **2017,** *50* (8), 085005.
5. Wu, K.; Su, D.; Liu, J.; Wang, J.-P., Estimating saturation magnetization of superparamagnetic nanoparticles in liquid phase. *Journal of Magnetism and Magnetic Materials* **2019,** *471*, 394-399.